\documentclass[11pt]{article}
\usepackage{multicol}
\usepackage{graphicx}
\usepackage{amssymb,bm,mathrsfs,bbm,amscd}
\usepackage[tbtags]{amsmath}
\usepackage{lastpage}
\usepackage{indentfirst}
\date{ }
\title{ Simulation Study of Laser Plasma Accelerator Via Vorpal }

\author{ Xiongwei Zhu \\
{\it Institute of High Energy Physics,  P.O.Box 918, }\\
{\it Chinese Academy of Sciences, Beijing 100049 } }

\begin{document}

\maketitle

\begin{abstract}
In this paper, we use PIC code Vorpal to do the extensive simulation about the laser plasma accelerator in the linear, quasilinear and nonlinear regime respectively. We design the $ ~100 MeV $ or so laser plasma accelerator ( LPA ) via Vorpal simulation. Finally, we discuss the application of the designed LPA in the compact light source field.
\end{abstract}



\section{Introduction }
Since the first proposal of laser wakefield accelerator by Tajima and Dawson\cite{a,b}, the laser plasma accelerator has gone through over thirty years\cite{aa}. In laser plasma accelerator, the longitudinal Langmuir wave is excited by the driving laser which passes through the plasma.  The perturbated electron density excites the plasma wave potential which forms the accelerating field as in conventional accelerator\cite{bb}. With the appearance of the quasi-monoenergetic electron beam from laser plasma accelerator ( LPA ), there begins the new hot advanced accelerator concept expedition to the laser plasma accelerator. The produced beam quality of LPA depends on the injection method\cite{c,d,e,f,g,a1,b1,c1}. The present main injection methods are the bubble mechanism\cite{c,d,e}, the colliding optical pulse injection\cite{f,g}, the controlling gradient method\cite{a1,b1,c1,d1,e1,f1,g1}, and the external injection method\cite{h1}. After the great achievement of the quasi-monoenergytical electron bunch during the past ten years, it seems that we come to the difficult situation to go further to produce the high quality electron beam. We need the new breakthrough coming out of our efforts.

\section{Vorpal code }
 Now, PIC code technique\cite{a2} is widely used in the accelerator simulation field.
 In the past, we have used OOPIC\cite{a3} to do the extensive simulation about the laser plasma accelerator\cite{d3,e3,f3}. Vorpal\cite{b3} is a fully object-oriented code including PIC and fluid simulation, and is widely used in the accelerator labs. We can use Vorpal to simulate the laser plasma accelerator in 1D, 2D, and 3D geometry. Vorpal is a good simulator for the plasma based accelerator simulation. My impression is that Vorpal is enough for the accelerator simulationist. According to our experience, OOPIC and Vorpal have become the main simulators of our simulations of the plasma based accelerator. In this paper, we mainly discuss the
 simulation of LPA via Vorpal.

\section{Simulation of Laser Plasma Accelerator via Vorpal}
In reference\cite{d3}, we give the simply instinctive theory of the working regimes about LPA. According to the laser strength, the laser plasma accelerator can be defined into three regimes: the linear regime, the quasilinear regime, and the nonlinear regime. In case of the linear regime or the linear theory, $ a_0 \ll 1$, the driving laser pulse will be guided well, and
the excited plasma wave wakefield has fine consine structure of a few of cycles. While, in case of the nonlinear case or the blow-out regime, $ a_0 \gg 1$. The bubble mechanism works in this regime and can produce the quasi-monoenergytical electron beam. This kind of injection method work stably, but the wakefield wave form is not so perfect to go further to obtain the electron beam of the more better quality. In case of the quasilinear regime, $ a_0 \sim 1$. This is the mid way to get the high quality electron beam. The quasilinear regime has the partial advantages of both the linear regime and the nonlinear regime. The bubble mechanism, the colliding optical pulse injection, the gradient control injection and the external injection method can work in this regime. In the
past ten years, the bubble mechanism has made great breakthrough in the expedition to the
advanced accelerator concept. The relative energy spread of LPA with the bubble mechanism
now stay at the level of few or ten percent. It may be better to return to the case of relatively small $ a_0 $ to have the plasma wave wakefield of the perfect wave form/shape.

In order to test the incorporated code Vorpal, we do the extensive one dimension, two dimension and three dimension simulations via Vorpal to validate the instinctive theory of the three kinds of regime mentioned above.
We use window platform to run the code vorpal. The basic laser parameters are the wavelength of $ 800 nm $, the pulse length of $ 1.6 \mu m $ . As for the laser strength, we set $a_0= 0.1, 0.5, 0.8, 0.9, 1.0, 1.1, 1.2, 4, 10 $ respectively. So we can simulate the linear regime, the quasilinear regime and the nonlinear regime or the bubble regime. The basic plasma parameter is the plasma density of $ 6 \times 10^{25} m^{-3} $. The simulation results are shown in the Figure 1-18. Figure 1-9 show the evolutions of the electric field. While Figure 10-18 show the evolutions of the plasma electrons. The simulation result agree the instinctive theory result basically.

\begin{center}
\includegraphics[width=14cm]{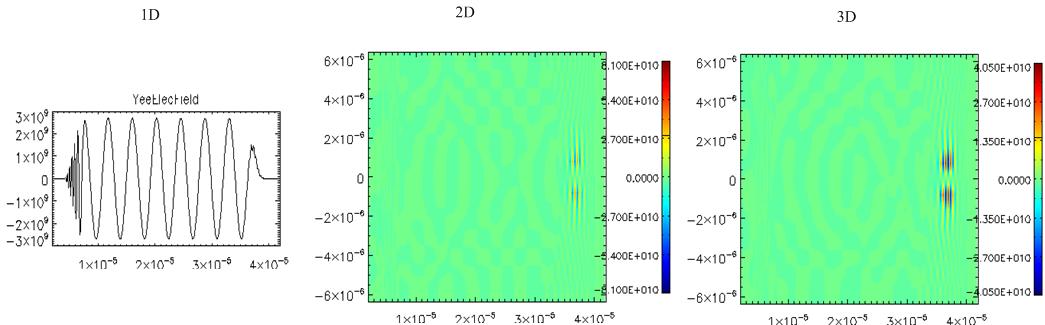}

{ Figure 1. the longitudinal electric field ( $a_0 = 0.1 $ ) }
\end{center}

\begin{center}
\includegraphics[width=14cm]{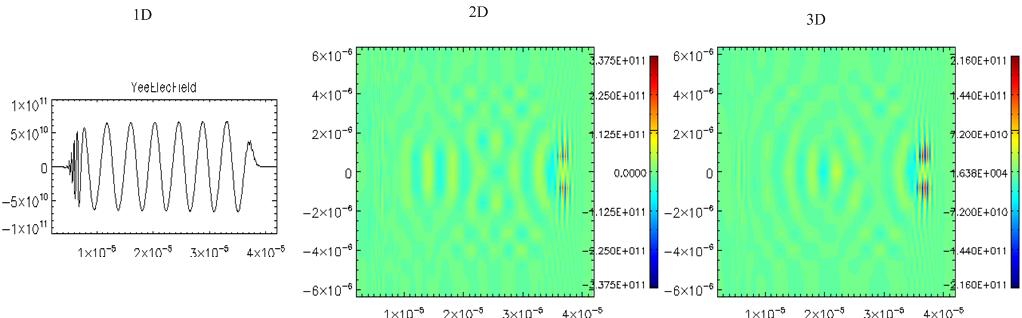}

{ Figure 2. the longitudinal electric field ( $a_0 = 0.5 $ ) }
\end{center}

\begin{center}
\includegraphics[width=14cm]{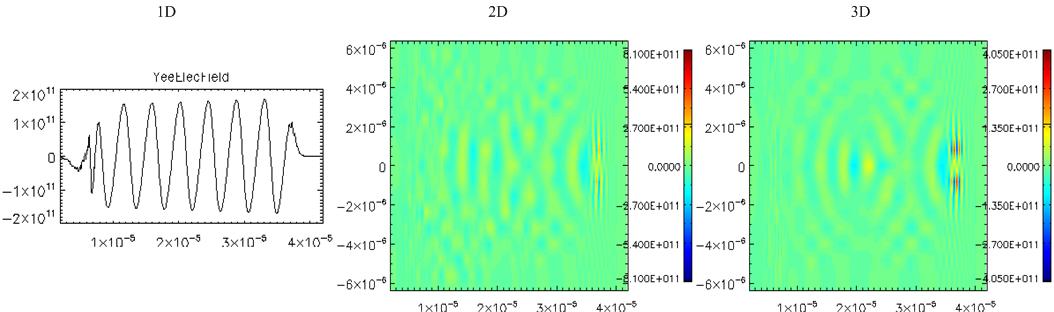}

{ Figure 3. the longitudinal electric field ( $a_0 = 0.8 $ ) }
\end{center}

\begin{center}
\includegraphics[width=14cm]{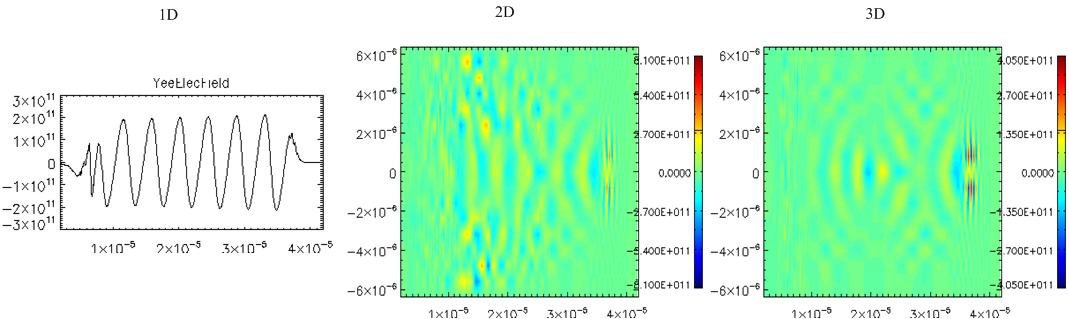}

{ Figure 4. the longitudinal electric field ( $a_0 = 0.9 $ ) }
\end{center}

\begin{center}
\includegraphics[width=14cm]{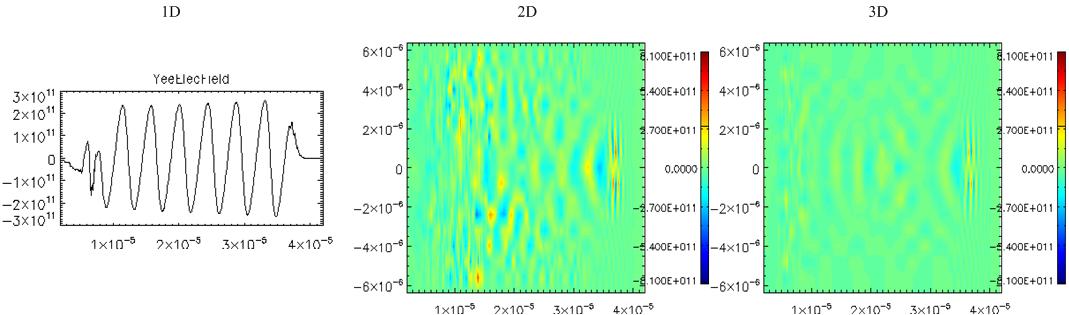}

{ Figure 5. the longitudinal electric field ( $a_0 = 1.0 $ ) }
\end{center}

\begin{center}
\includegraphics[width=14cm]{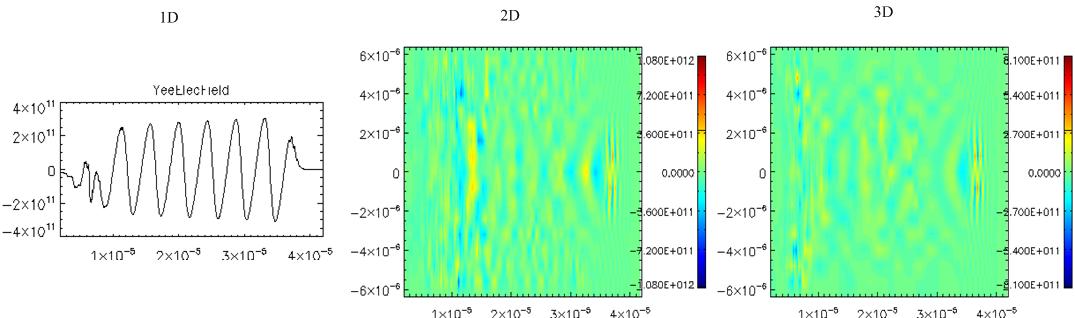}

{ Figure 6. the longitudinal electric field ( $a_0 = 1.1 $ ) }
\end{center}

\begin{center}
\includegraphics[width=14cm]{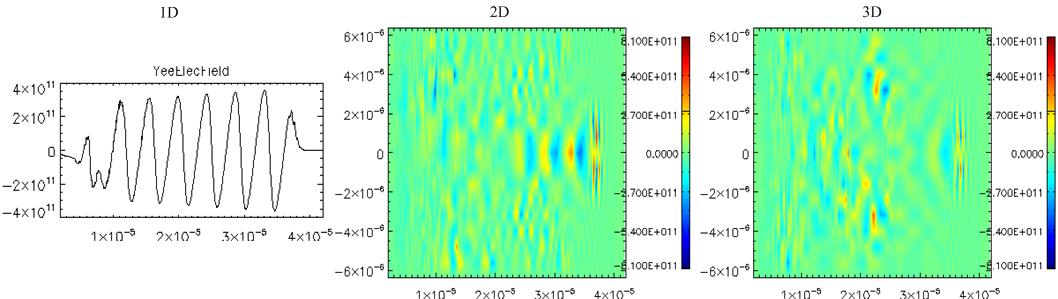}

{ Figure 7. the longitudinal electric field ( $a_0 = 1.2 $ ) }
\end{center}

\begin{center}
\includegraphics[width=14cm]{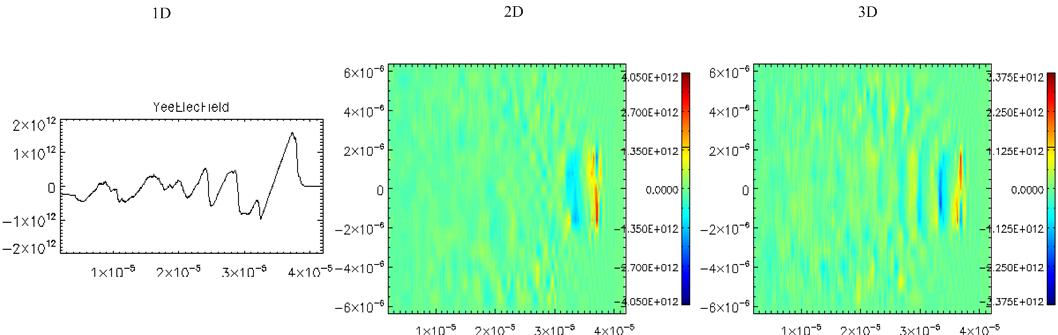}

{ Figure 8. the longitudinal electric field ( $a_0 = 4.0 $ ) }
\end{center}

\begin{center}
\includegraphics[width=14cm]{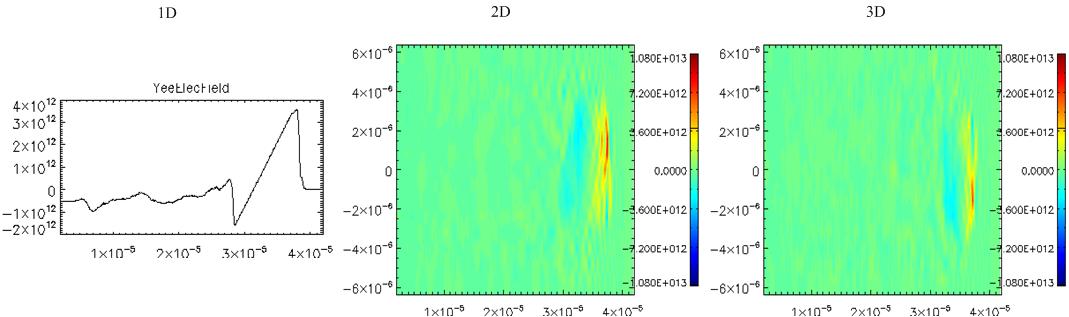}

{ Figure 9. the longitudinal electric field ( $a_0 = 10.0 $ ) }
\end{center}

\begin{center}
\includegraphics[width=14cm]{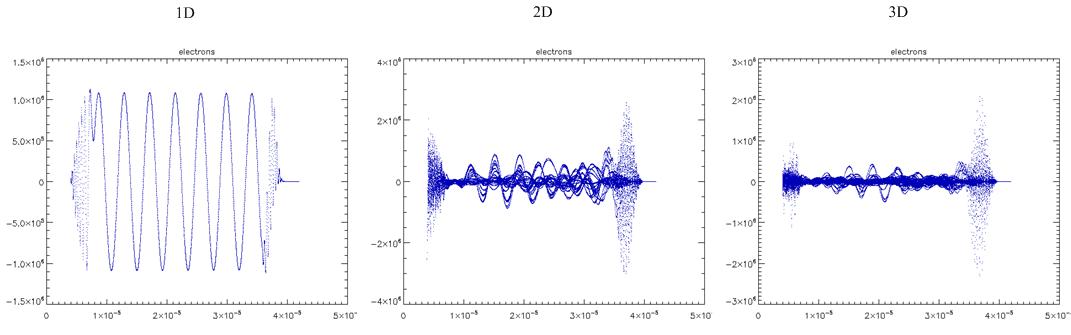}

{ Figure 10. the longitudinal phase space ( $a_0 = 0.1 $ ) }
\end{center}
\begin{center}
\includegraphics[width=14cm]{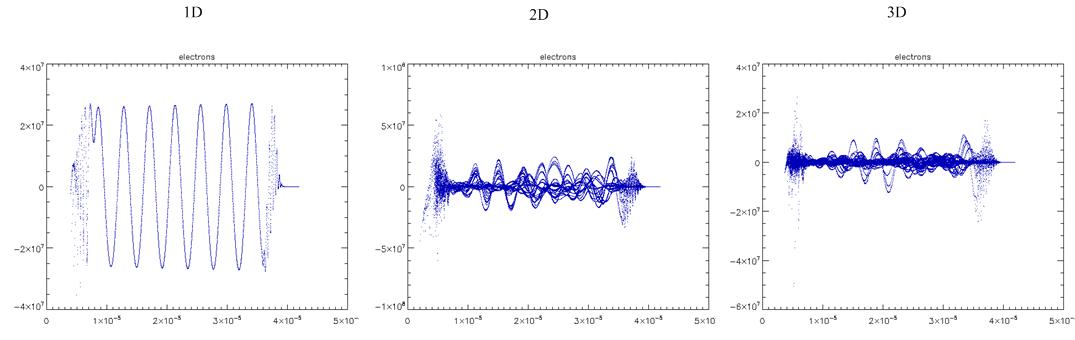}

{ Figure 11. the longitudinal phase space ( $a_0 = 0.5 $ ) }
\end{center}
\begin{center}
\includegraphics[width=14cm]{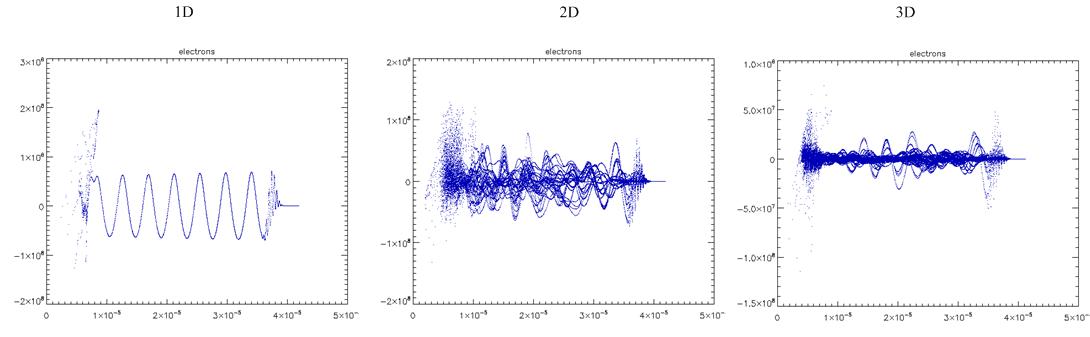}

{ Figure 12. the longitudinal phase space ( $a_0 = 0.8 $ ) }
\end{center}
\begin{center}
\includegraphics[width=14cm]{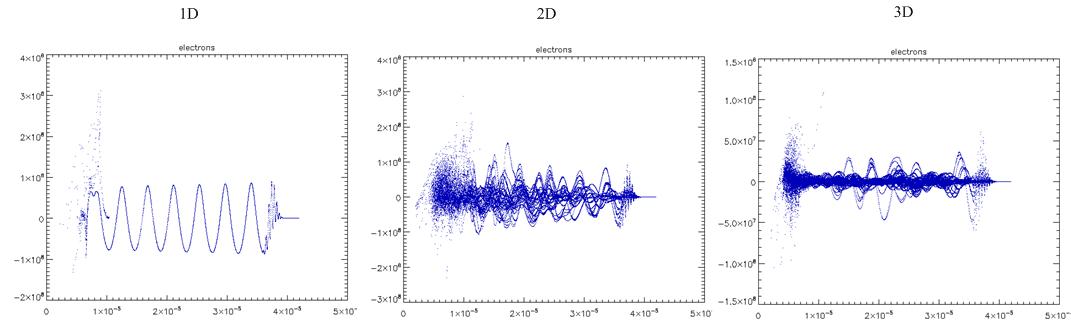}

{ Figure 13. the longitudinal phase space ( $a_0 = 0.9 $ ) }
\end{center}
\begin{center}
\includegraphics[width=14cm]{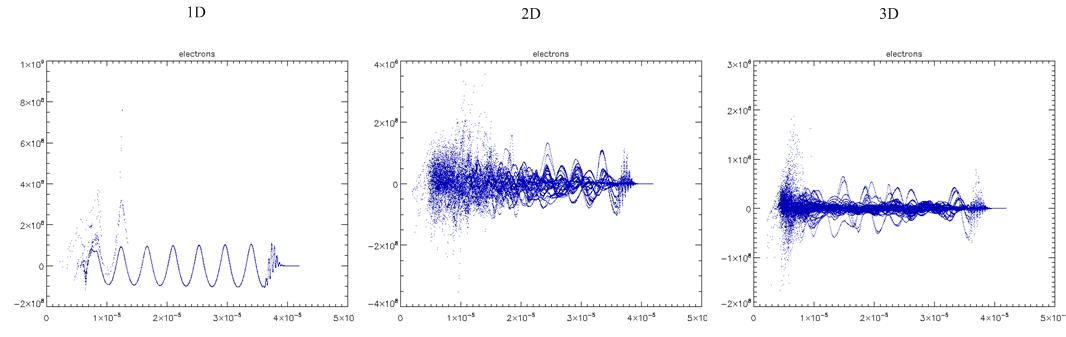}

{ Figure 14. the longitudinal phase space ( $a_0 = 1.0 $ ) }
\end{center}
\begin{center}
\includegraphics[width=14cm]{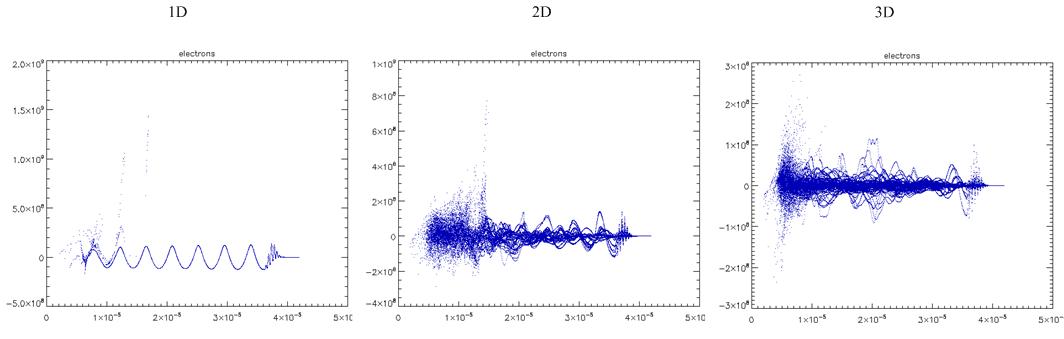}

{ Figure 15. the longitudinal phase space ( $a_0 = 1.1 $ ) }
\end{center}
\begin{center}
\includegraphics[width=14cm]{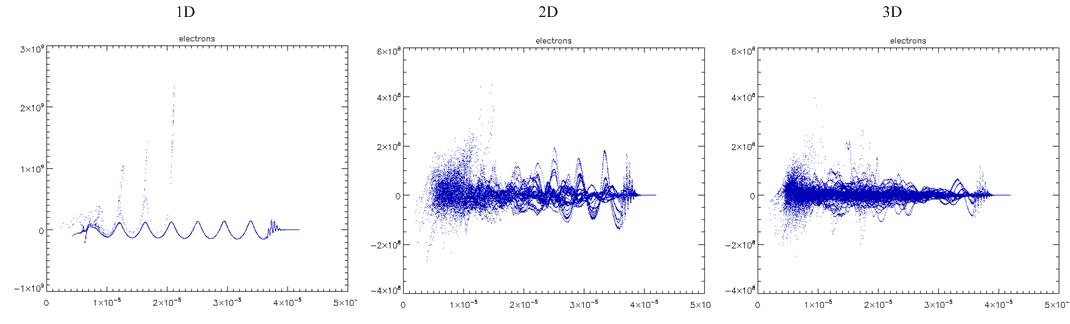}

{ Figure 16. the longitudinal phase space ( $a_0 = 1.2 $ ) }
\end{center}
\begin{center}
\includegraphics[width=14cm]{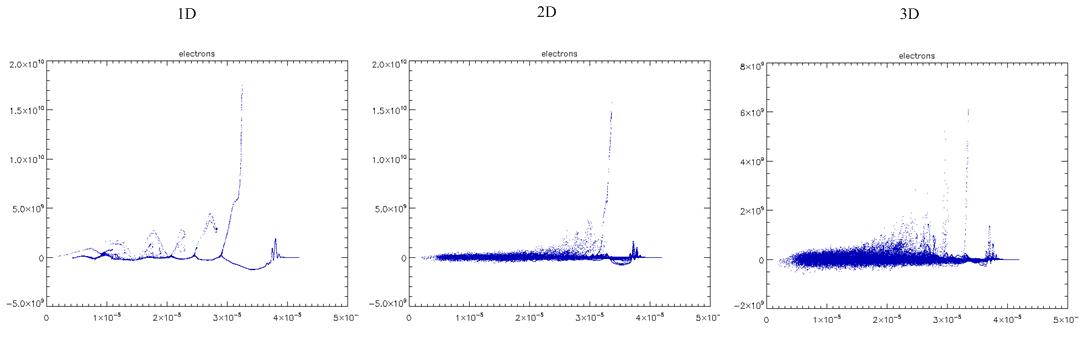}

{ Figure 17. the longitudinal phase space ( $a_0 = 4.0 $ ) }
\end{center}
\begin{center}
\includegraphics[width=14cm]{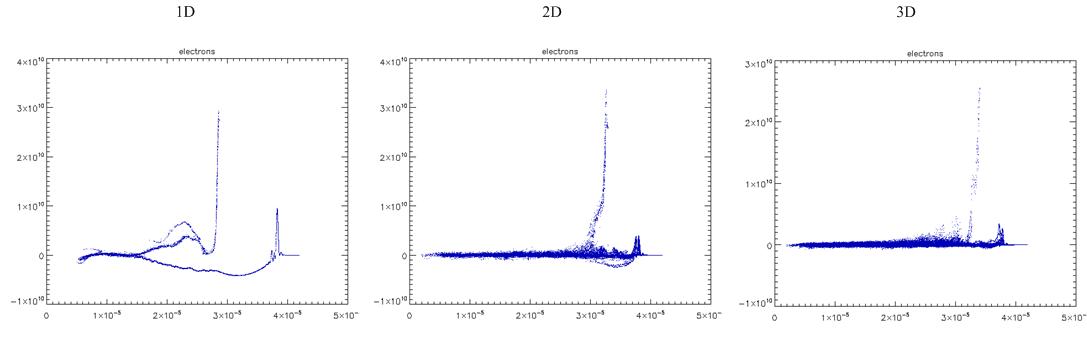}

{ Figure 18. the longitudinal phase space ( $a_0 = 10.0 $ ) }
\end{center}

\section{Design of a single cell LPA }
A $ 100 MeV $ stage laser plasma accelerator design simulation is presented. The simulation area is $80 \mu m \times 25.6 \mu m$, also the moving window technique is used. The used simulation PIC cell number is $ 160 \times 32 $. The plasma density is $ 6 \times 10^{19} cm^{-3} $. The laser plasma interaction works in the bubble mechanism. The laser strength is set to be $ a_0 = 15 $, the dephase distance and the depletion distance are estimated to be several hundreds of micrometers roughly £¬ while the Rayleigh length is just $ 20 \mu m $ or so, but the laser power is far bigger than the critical laser power, so there exists the relativistical and ponderomotive channel optical guiding phenomenon. the simulated interaction range is less than the dephase and the depletion distance, the simulation distance is several times of the Rayleigh length. The estimated maximal energy agrees with the simulation result. Figure 19-21 summary the design simulation result.

\begin{center}
\includegraphics[width=14cm]{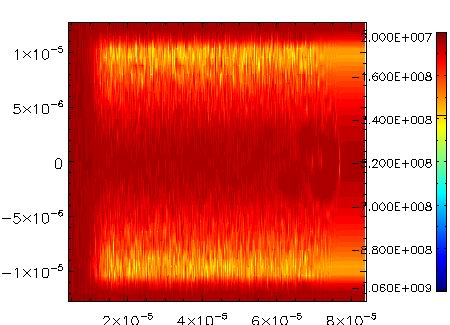}

{ Figure 19. the plasma density }
\end{center}

\begin{center}
\includegraphics[width=14cm]{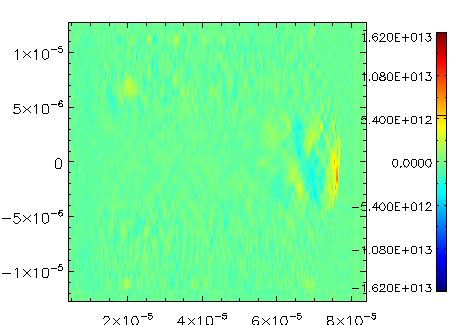}

{ Figure 20. the longitudinal electricfield ( $ a_0 = 15 $ ) }
\end{center}

\begin{center}
\includegraphics[width=14cm]{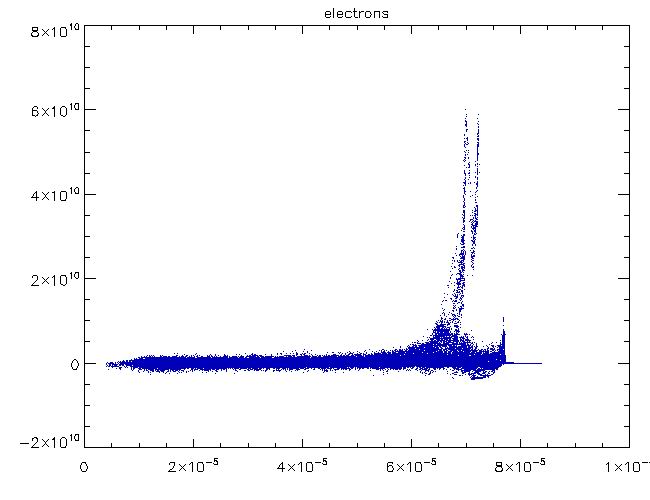}

{ Figure 21. the logitudinal phase space ( $ a_0 = 15 $ ) }
\end{center}

\section { Design for Thomson backscattering light source }
The electron beam drived x-ray light source can be defined into two classes: coherent light source and incoherent light source. In case of high energy electron beam, free electron laser is the coherent light source, while the synchrotron radiation is the incoherent light source. The low energy electron beam can also produce the X-ray radiation. One of the methods for the low energy electron beam produced x-ray radiation is the Thomson scattering\cite{c3}.
When a laser beam ( with the frequency $ \omega_0 $ )interacts the electron beam at an angle $ \phi $, the upshift frequency of the laser as

\begin{equation}
\omega = \frac{2\gamma^2 ( 1 - cos\phi )}{1 + \frac{a_0^2}{2}} \omega_0.
\end{equation}

where $ \gamma $ is the lorentz factor, $ a_0 $ is the normalized laser strength. Using this mechanism, we can design the experiment scheme for x-ray radiation facility based on the 100 MeV electron beam from the laser plasma accelerator. Figure 22
shows the basic setup of the compact x-ray production facility.

\begin{center}
\includegraphics[width=14cm]{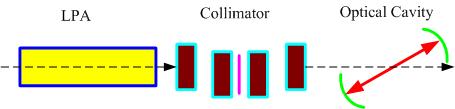}

{ Figure 22. the scheme for the x-ray experiment }
\end{center}

We can use the collimator to obtain the electron beam with the required energy spread. The typical parameters of the output electron bunch can be the charge of $ 10 pC $, the energy of $ 100 MeV $, the energy spread of $ 5 \% $, and the normalized emittance of $ 0.1 mm mrad $ respectively. We have used the laser beam of the wavelength of $ 3 \mu m $ to do some laser electron beam scattering calculations. The output photon energy can be as high as $ 50 KeV $.

\end{document}